\documentclass[12pt]{iopart}
\usepackage{color}
\usepackage{graphicx}
\usepackage{epsfig}
\usepackage{verbatim}
\begin{document}
\title{Lamb Shift of Energy Levels in Quantum Rings}

\author{G Yu Kryuchkyan$^{1,2}$, O Kyriienko$^{3,4}$ and I A Shelykh$^{3,4}$}

\address{$^1$Yerevan State University, Centre of Quantum Technologies and New Materials, Alex Manoogian 1, 0025, Yerevan, Armenia}
\address{$^2$Institute for Physical Researches, National Academy of Sciences, Ashtarak-2, 0203, Ashtarak, Armenia}
\address{$^3$Science Institute, University of Iceland, Dunhagi 3, IS-107, Reykjavik, Iceland}
\address{$^4$Division of Physics and Applied Physics, Nanyang Technological University 637371, Singapore}

\begin{abstract}
We study the vacuum radiative corrections to energy levels of a confined electron in quantum rings. The calculations are provided for the Lamb shift of energy levels in low-momentum region of virtual photons and for both one-dimensional and two-dimensional quantum rings. We show that contrary to the well known case of a hydrogen atom the value of the Lamb shift increases with the magnetic momentum quantum number \emph{m}. We also investigate the dependence of the Lamb shift on magnetic flux piercing the ring and demonstrate a presence of magnetic-flux-dependent oscillations. For one-dimensional ring the value of the shift strongly depends on the radius of the ring. It is small for semiconductor rings but can attain measurable quantities in natural organic ring-shape molecules, such as benzene, cycloalcanes and porphyrins.
\end{abstract}

%
%

\section{Introduction}
Quantum electrodynamics (QED) is the most accurate theory known so far to investigate the fundamental processes of the matter-light interactions \cite{Itzykson,Mohr2000}. QED predicts the possibility to observe effects related to vacuum fluctuations, i.e. creation and absorbtion of virtual quanta of electromagnetic field referred as virtual photons. The most notable example is the Lamb radiative energy shift that was first observed in hydrogen atoms \cite{Lamb1947}. Among the other QED effects one can note the appearance of Casimir forces between parallel conducting plates and phenomena of vacuum polarization leading to photon-photon scattering in vacuum due to creation of virtual electron-positron pair. 

Since the original calculations of the value of the Lamb shift by Bethe and later by Feynman and others, the theory of radiative corrections has been worked out up to a very high precision. The Lamb radiative shift  since then has been experimentally investigated for various atomic systems including muonic atoms that provide an ultra-high precision test of the QED \cite{Hansch, Pohl2013}. The high accuracy of QED predictions for the precise spectroscopy of simple atomic systems allowed the accurate measurements of fundamental physical constants including the Rydberg constant $R_{\infty}$ from the hydrogen spectrum, $\alpha$ from the helium fine structure and the electron mass $m_e$ from the $g$ factor of hydrogen-like ions. The vacuum QED effects in the extremely strong atomic fields has also been measured in experiments with highly charged few-electron ions \cite{Mohr1998,Beier2000}. Besides, cavity QED allows to study the vacuum radiative shift in interaction of atoms with a single mode electromagnetic field  \cite{Heinzen1987,Brune1994,Marrocco1998,Haroche} and investigate the vacuum Rabi splitting in a system consisting of a single quantum dot placed into optical semiconductor cavity \cite{Khitrova2006,Schoelkopf2008}. The modification of the Lamb shift  in the presence of external strong laser fields  was also investigated \cite{Kryuchkov1982,Jentschura2003,Evers2004,Kryuchkyan2007,Kryuchkyan2009-1,Kryuchkyan2009-2}. 

Up to now, the study of radiation shifts was mostly restricted to the systems consisting of a single atom. However, the advances in the engineering of micro- and nanoelectromechanical systems, optical microcavities, superconducting circuits and artificial atoms made actual the problem of the investigation of quantum vacuum effects in these nanodevices. In this context, the Lamb shift has been observed in a superconducting electronic circuit in the form of a superconducting Cooper pair box in a transmission-line resonator \cite{Fragner2008}. It has been also shown that a superconducting qubit strongly coupled to a non-linear resonator can act as a probe of quantum fluctuations of the intra-resonator field. Theoretical and experimental results have been presented in Refs. \cite{Ong2013,Ong2011,Boissonneault2012}.

In addition to these important results it would be interesting to study the vacuum radiative corrections in other types of the nano-scale systems. The natural candidates here are Aharonov-Bohm quantum rings, where the spectrum of the discrete states can be easily tuned by application of the external magnetic field due to the Aharonov-Bohm effect \cite{Ehrenberg1949,Aharonov1959}. The investigation of QED effects in ring-based structures has already started. In particular, it was theoretically demonstrated that in the chiral optical resonators the ground state of electron system in the ring can be associated to non-zero angular momentum \cite{Kibis2013,Kibis2011a,Kibis2012}. Additionally, an interaction of electron with circularly polarized photons has shown to modify charge and spin flow in a quantum ring \cite{Arnold2013,Arnold2014a,Arnold2014b}. In this paper we make further contribution to this domain and present the calculation of the vacuum radiative Lamb shifts for 1D and 2D quantum rings placed in the external magnetic field (Fig. 1) calculating the self-energy of confined electron including also mass-renormalization procedure. As predicted by QED such bound-state self-energy part is the dominant radiative correction in hydrogen-like systems and gives 98$\%$ of the ground-state Lamb shift in atomic hydrogen \cite{Jentschura1999}.

We demonstrate, that the Lamb shift  is minimal for the state with minimal value of $m+f$, where $m$ is a value of the electrons angular momentum and $f$ is a magnetic flux piercing the ring. This is qualitatively different to the case of a hydrogen atom, where the Lamb shift is maximal for $s$-states. Besides, we demonstrate that the value of the Lamb shift of momentum levels reveals periodical dependence on $f$  specific for  Aharonov-Bohm system.

For 2D quantum ring  the energy spectrum consists of the discrete energy levels due to radial motion with the radial quantum numbers $n= 1, 2, ...$, and rotation motion with the quantum numbers $m=0, \pm 1,  ...$. In this case we present calculations in so-called Bethe logarithmic approximation in analogy to consideration of real atoms. This allows us to obtain the general approximate result for the Lamb shift without specification of the confining potential. 
\begin{figure}
\centering
\includegraphics[scale=0.4]{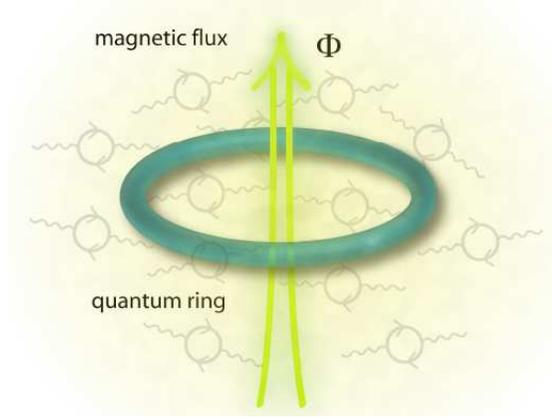}
\caption{The sketch of the system depicting an electron in a quantum ring interacting with electromagnetic vacuum fluctuations, and additionally subjected to supplemental magnetic field in the Aharonov-Bohm geometry.}
\label{Fig1}
\end{figure}

\section{The interaction Hamiltonian}

Let us consider the system shown schematically in Fig. 1. The Hamiltonian of an electron in a quantum ring interacting with radiation field reads as
\begin{equation}
H=H_{0}+H_{R}+H_{int},
\label{energy}
\end{equation}
where $H_{0}$ and $H_{R}$ are the free Hamiltonians of the quantum ring and the radiation field, and
\begin{equation*}
H_{int}=\int{\vec{A(r)}\vec{j(r)}d^{3}r}
\end{equation*}
describes the interaction of a confined electron with the radiation field, where $\vec{j}$ is the electron current operator and $\vec{A}$ is the vector potential. We use the Furry representation for the vector state of the system with the Hamiltonian (\ref{energy})
\begin{equation}
|\psi(t)\rangle=U(t)e^{-\frac{i}{\hbar} H_{R}t}|\phi(t)\rangle,
\end{equation}
where
\begin{equation}
U(t)=e^{-\frac{i}{\hbar}H_{0}t}.
\end{equation}
In this representation the dynamic equation for the state $|\phi(t)\rangle$ reads
\begin{equation}
i\hbar\frac{\partial}{\partial t}|\phi(t)\rangle =H_{1}(t)|\phi(t)\rangle ,
\label{evol}
\end{equation}
where
\begin{equation}
H_{1}(t)=U^{-1}(t)e^{\frac{i}{\hbar}H_{R}t}H_{int}e^{-\frac{i}{\hbar}H_{R}t}U(t)=\int \vec{A}(r,t)\vec{j}(r,t)dV,
\label{H1}
\end{equation}
and the current operator in the Furry representation reads
\begin{equation}
\vec{j}(r,t)=U^{-1}(t)\vec{j}(r)U(t).
\end{equation}
The system displays cylindrical symmetry thus we use the circular polarizations, $\hat{e}_{+}=\hat{e}_{x}+i\hat{e}_{y},
\hat{e}_{-}=\hat{e}_{x}-i\hat{e}_{y}$
and the field operator can be written in the following form:
\begin{equation}
\vec{A}(r,t)=\sqrt{\frac{\hbar c}{(2\pi)^{3}}}\sum_{\lambda=+,-}\frac{d^{3}k}{\sqrt{2\omega_{k}}}\left[\hat{e}_{\lambda}(k)A_{\lambda}(k)\right],
\end{equation}
where
\begin{eqnarray} 
A_{+}(k)=a_{+}(k)e^{-i(\omega_{k}t-\vec{k}\vec{r})}+a_{-}^{\dagger}(k)e^{i(\omega_{k}t-\vec{k}\vec{r})}\nonumber, \\
A_{-}(k)=a_{-}(k)e^{-i(\omega_{k}t-\vec{k}\vec{r})}+a_{+}^{\dagger}(k)e^{i(\omega_{k}t-\vec{k}\vec{r})}.
\end{eqnarray}
Here: $a_{\lambda}^{\dagger}(k)$ corresponds to photon creation operator where sub-script describes the state of
 photon polarization, while $a_{\lambda}(k)$ denotes the photon annihilation operator.

The current operator $\vec{j}=e\vec{v}$, where $\vec{v}$ is the velocity operator of confined electron, in the cylindrical coordinates can be written in two-component form as
\begin{equation}
\vec{j}=e(\hat{\rho}v_{\rho}+\hat{\varphi} v_{\varphi})
\end{equation}
where $\hat{\rho},\hat{\varphi}$ are unity vectors, and the interaction Hamiltonian reads:
\begin{equation}
\fl H_{int}=e\sqrt{\frac{\hbar c}{(2\pi)^{3}}}\int\frac{d^{3}k}{\sqrt{2\omega_{k}}}\bigg[\left(A_{+}(k)e^{i\varphi}+A_{-}(k)e^{-i\varphi}\right)v_{\rho}+i\left(A_{+}(k)e^{i\varphi}-A_{-}(k)e^{-i\varphi}\right)v_{\varphi}\bigg]dV.
\end{equation}

\section{Matrix elements of the radiative transitions}

In this section we derive the matrix elements of transitions between states of confined electron for 1D and 2D models produced by the term of electron-radiation field interaction.

\subsection{The case of 1D quantum ring}

The Hamiltonian of an electron confined in 1D infinitely narrow quantum ring depends only on the polar angle $\varphi$. We consider a general case of the quantum ring pierced by the magnetic flux $\Phi$. The vector-potential is chosen as $\vec{A}=\frac{\Phi}{2\pi R}\hat{\varphi}$, where R is the radius of a ring, and electron momentum operator is $\vec{P}=-i\hbar\hat{\varphi}\frac{1}{R}\frac{\partial}{\partial\varphi}$. The corresponding Hamiltonian is given by the expression
\begin{equation}
H_{0}(\varphi)=\frac{1}{2m_{e}}\left(\vec{P}-\frac{e}{c}\vec{A}\right)^{2}=
-\frac{\hbar^{2}}{2m_{e}R^{2}}\left(\frac{\partial}{\partial\varphi}+if\right)^{2},
\label{mhamilton}
\end{equation}
where $m_{e}$ is the electron mass (or the effective mass), $f=\frac{\Phi}{\Phi_{0}}$ is the number of flux quanta piercing the ring, $\Phi_{0}=h/e$. The $2\pi$-periodic eigenfunctions and the energy eigenvalues of the system are
\begin{eqnarray}
\psi_{m}(\varphi)=\frac{1}{\sqrt2\pi}e^{im\varphi},~~~~~~
\varepsilon_{m}=\varepsilon_{0}(m+f)^{2},
\end{eqnarray}
where $\varepsilon_{0}=\hbar^{2}/2m_{e}R^{2}$.

At first, we calculate the matrix elements of the radiative transitions $\langle n|(\hat{e}_{\pm}\vec{v})|m\rangle$ between states $|m\rangle$ and $|n\rangle$ of the rotation motion with angular momentum quantum numbers $m=0$, $\pm 1$, $\pm 2$, ... . For this goal we consider the Heisenberg equation for the radial vector $\vec{r}(t)=U^{-1}(t)\vec{r}U(t)$, where $\vec{r}=\hat{\rho}\rho$, in the standard form
\begin{equation}
\vec{v}=\frac{d}{dt}\vec{r}=\frac{i}{\hbar}\left[H_{0}(\varphi),\vec{r}\right].
\end{equation}
Using Eq. (\ref{mhamilton}) and the expressions for the basis vectors
\begin{eqnarray}
\frac{\partial\hat{\rho}}{\partial\varphi}=\hat{\varphi},~~~~~~
\frac{\partial\hat{\varphi}}{\partial\varphi}=-\hat{\rho},
\end{eqnarray}
we calculate the velocity operator in the following two-component form
\begin{equation}
\vec{v}=\frac{i\hbar}{2m_{e}R^{2}}\left[\hat{\rho}-2\hat{\varphi}\left(\frac{\partial}{\partial\varphi}+if\right)\right].
\end{equation}
Then, the formulas $(\hat{e}_{\pm}\hat{\rho})=e^{\pm i\varphi}$, $(\hat{e}_{\pm}\hat{\varphi})=\pm ie^{\pm i\varphi}$ are used, and integration on the azimuthal angle can be performed using relation
\begin{equation}
\frac{1}{2\pi}\int e^{i(n-m)\varphi}d\varphi=\delta_{n,m},
\end{equation}
where $\delta_{n,m}$ is Kronecker delta function.
Finally, we obtain the radiative transitions in the following forms
\begin{equation}
\langle n|(\vec{e}_{\pm}\vec{v})|m\rangle=\frac{i\hbar}{2m_{e}R}[1\pm 2(m+f)]\delta_{n, m\pm 1} \label{matrix1}.
\end{equation}

\subsection{The case of 2D quantum ring}

For 2D quantum ring the Hamiltonian  involves also the radial dependence and reads as
\begin{equation}
H_{0}=-\frac{\hbar^{2}}{2m_{e}}\left(\frac{\partial^{2}}{\partial\rho^{2}}+\frac{1}{\rho}\frac{\partial}{\partial\rho}\right)
-\frac{\hbar^{2}}{2m_{e}\rho^{2}}\frac{\partial^{2}}{\partial\varphi^{2}}+V(\rho),
\label{Ham2D}
\end{equation}
where $V(\rho)$ is the confining potential. The energy spectrum of $H_{0}$ consists of the discrete energy levels $E_{N}=E_{n, m}$ due to radial motion with the  radial quantum numbers $n= 1, 2, ...,$ and rotation motion with the quantum numbers $m=0, \pm 1, \pm 2, ...$. The joint states are denoted as $|N\rangle=|R_{n, m}\rangle|m\rangle$.

The velocity operator $\vec{v}(t)=U^{-1}(t)\vec{v}U(t)$ in this case is calculated in the following form
\begin{equation}
\vec{v}(t)=\frac{i}{\hbar}[H_{0}, \vec{r}]=-\frac{i\hbar}{m_{e}}\left(\hat{\rho}\frac{\partial}{\partial\rho}+\hat{\varphi}\frac{1}{\rho}\frac{\partial}{\partial\varphi}\right)
\end{equation}
that is different from the analogous result for $1D$ model.
Then, by using this formula the matrix elements of the radiative transitions can be calculated as
\begin{equation}
\langle m'|\langle R_{n', m'}|\left(\hat{e}_{\pm}\vec{v}\right)|R_{n,m}\rangle|m\rangle =\frac{i}{\hbar}\left(E_{n', m'}-E_{n, m}\right)R_{n', m'; n, m}\delta_{m', m\pm 1},
\label{vtranz}
\end{equation}
where
\begin{eqnarray}
\label{Rnmnm}
R_{n', m'; n, m}=\int_{0}^{\infty}R^{*}_{n', m'}R_{n, m}\rho^{2}d\rho
\end{eqnarray}
and the normalization condition for the radial wave functions reads as
\begin{equation}
\int_{0}^{\infty}R^{*}_{n', m'}R_{n, m}\rho d\rho=\delta_{n', n}\delta_{m', m}.
\end{equation}

\section{Radiative shifts of the energy levels}

In this section we derive the general expression for the radiative shift of confined electron using the expressions for the transition matrix elements obtained in the previous section. The time evolution of the system vector state due to interaction with the radiation field in the Furry picture is given by Eq. (\ref{evol}). The formal solution of this equation can be written as
\begin{eqnarray}
|\phi(t_2)\rangle = U_{F}(t_2,t_1)|\phi(t_1)\rangle,\\
U_{F}(t_2,t_1) = T \exp \left( -i \int_{t_1}^{t_2} d\tau H_1(\tau) \right),
\end{eqnarray}
through the time-evolution matrix $U_{F}(t_2,t_1)$,  where $T$ denotes the time-ordering symbol. If 2D ring is considered in the Furry picture, the confinement potential $V(\rho)$ is included into free Hamiltonian.

The radiative shift of the $E_{n,m}$ energy level for $|N\rangle =|\Psi_{n,m}(t)\rangle= |R_{n,m}\rangle |m\rangle$  state of an electron confined in the ring is expressed through time-evolution matrix by using the Tomonaga-Schwinger equation
\begin{equation}
e^{-i\Delta E_{n,m}(t-t_0)}=\frac{\langle 0| \langle N| U_{F}(t,t_0)|N\rangle |0\rangle}{\langle 0| \langle N_0| U_{F}(t,t_0)|N_0 \rangle |0\rangle}.
\end{equation}
Here $|N_0\rangle$ and $|0\rangle$ are the vacuum states of an electron and a photon, respectively.

\begin{figure}
\centering
\includegraphics[width=0.5\linewidth]{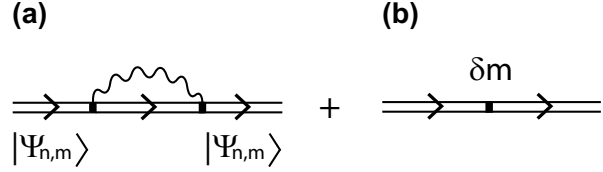}
\caption{(a) Diagram of the self-energy part and (b) diagram of the mass renormalization. Double lines denote the states and propagation function of the confined electron. Wavy line denotes the photon propagation function.}
\label{Fig2}
\end{figure}

In the second-order of the perturbation theory, the following expression for radiative shifts of $E_{n,m}$ energy levels may be easily obtained:
\begin{equation}
\label{shiftTS2}
(t-t_0) \Delta E_{n,m} = -i\int_{t_0}^{t}d\tau_1 \int_{t_0}^{\tau_1}d\tau_2 \langle 0|\langle N|H_1(\tau_1)H_1(\tau_2)| N\rangle |0 \rangle .
\end{equation}
In this expression we neglected the effects of the vacuum polarization for confined electron. Thus, Eq. (\ref{shiftTS2}) represents only self-energy part of the radiative shift. In Fig. \ref{Fig2} we show, for completeness, corresponding Feynman diagrams for the self-energy and the electron mass renormalization.
Note that for the case of stationary confined potential integration over time in the formula (\ref{shiftTS2}) is easily performed leading to the factor $(t-t_0)$ and hence to the definite result for $\Delta E_{n,m}$.

Typically, the calculation of the Lamb shift for atomic systems was made by splitting the basic formula for radiative shift in two parts that correspond to electron-photon interaction in two spectral ranges of virtual photon \cite{Berestecki}. For low-momentum region,  up to some momentum of virtual photon $k$ of the order of $k_{max}\leq(\alpha Z)m_{e}$, the exact wave functions of confined electron including all orders of interaction with potential $V$ are used, however in non-relativistic approach. In high-momentum range $k>k_{max}$ the electron wave function can be used in the first approximation of perturbation theory on atomic potential, but in relativistic approach. In general, by adding two parts, the full Lamb shift does not depend on the cut-off parameter $k_{max}$.

The analogous approach is used in the case of quantum ring. In this area we restrict ourself to the calculation of the non-relativistic part of the Lamb shift. In this way, the standard calculations including also the electron mass renormalization procedure lead to the following result ($\hbar=c=1$):
\begin{equation}
\fl \Delta E_{n,m}=\frac{\alpha}{4\pi^{2}}\int_{0}^{k_{max}}\frac{d^{3}k}{{\omega_{k}} ^{2}}\sum_{N', \lambda=+,-}(E_{N'}-E_{N}) \frac{\langle N|({\hat{e}_{\lambda}}^{*}\vec{v})|N'\rangle\langle N'|(\hat{e}_{\lambda}\vec{v})|N\rangle}{\omega_{k}-(E_{N}-E_{N'})},
\label{Eshift}
\end{equation}
where $\alpha$ is fine structure constant and $k_{max}$ is the cut-off parameter and the matrix elements of transitions are calculated in Sec. 3.

It is well known that the mass renormalization is usually realized by adding the term $\frac{\delta m}{2}\langle N|v^{2}|N\rangle$ described by Fig. \ref{Fig2}(b) for $\vec{j}\vec{A}$ version of electron-radiation field interaction. In this case the low-frequency part of the shift (\ref{Eshift}) contains only logarithmic divergence that is regularized by the cut-off parameter $k_{max}$. The analogous procedure can not be realized for the dipole version $\vec{E}\vec{d}$ of the interaction because in this case the formula (\ref{Eshift}) also contains additional divergences.  Note that Eq. (\ref{Eshift}) is valid in the long-wavelength approximation $kR\ll 1$ and hence the integration takes place until the energy of cutoff $k_{max}$. It is reasonable to take $k_{max}<\frac{1}{R}\approx\sqrt{m_{e}\varepsilon}$, where $\varepsilon$ is the characteristic energy of confined electron. On the other hand, $k_{max}$ can not be arbitrary small. It is easy to realize assuming that the propagator of confined electron differs from the propagator of free electron (see Fig. \ref{Fig2}). This consideration leads to the following inequalities $kp\approx k_{max}m_{e}\gg p^{2}-m_{e}^{2}\approx m_{e}\varepsilon$, where we introduced four-vector of the electron momentum $p_{\mu}=(m_{e}+\varepsilon,\overrightarrow{p})$. Thus, we obtain $\varepsilon\ll k_{max}\ll\sqrt{m_{e}\varepsilon}$.

Below we apply the general expression (\ref{Eshift}) to 1D and 2D models of quantum rings corresponding to the Hamiltonians (\ref{mhamilton}) and (\ref{Ham2D}).

\subsection{Radiative shift for 1D quantum ring}

In this subsection we calculate radiative shift for a one-dimensional  quantum ring with the vanishing width $d$ for which we can consider only azimuthal energy level structure. By using the matrix element (\ref{matrix1}) for the radiative shift $\Delta E_{m}(f)$ of the energy levels of a quantum ring pierced by the magnetic flux, $\varepsilon_{m}=\varepsilon_{0}(m+f)^{2}$,  ($\varepsilon_{0}=\hbar^{2}/2m_{e}R^{2}$), we obtain from the formula (\ref{Eshift})
\begin{equation}
\fl \Delta E_{m}(f)=\frac{\alpha{\varepsilon_{0}}^{2}}{2\pi m_{e}}\bigg[(1-2(m+f))^{3}\ln\frac{k_{max}}{\varepsilon_{0}|1-2(m+f)|} +(1+2(m+f))^{3}\ln\frac{k_{max}}{\varepsilon_{0}|1+2(m+f)|}\bigg],
\label{shift1D}
\end{equation}
where $k_{max}$ is the energy of cut-off, $\varepsilon_{0}\ll k_{max}\ll\sqrt{m_{e}\varepsilon_{0}}$.

The radiative shift (\ref{shift1D}) for the lowest energy state corresponding to minimal value $m+f=0$ is periodic in flux with the period $\Phi_{0}$ and reads
\begin{equation}
\label{shitf1Dfin}
\Delta E_{m}(f=-m)=\frac{\alpha{\varepsilon_{0}}^{2}}{\pi m_{e}}\ln\left(\frac{k_{max}}{\varepsilon_{0}}\right).
\end{equation}
We also note the symmetry of the Lamb shifts relatively to the sign of azimuthal quantum numbers, $\Delta E_{m}(f)=\Delta E_{-m}(-f)$.

We can rewrite equation (\ref{shift1D}) in the following form:
\begin{equation}
\Delta E_{m}(f)=[1+12(m+f)^2]\delta_{m}(f=-m) - \frac{\alpha{\varepsilon_{0}}^{2}}{2\pi m_{e}}\bigg[\Lambda_{-}^3 \ln|\Lambda_{-}| + \Lambda_{+}^3 \ln|\Lambda_{+}| \bigg],
\label{shift1D(2)}
\end{equation}
where $\Lambda_{\pm} = 2(m+f)\pm 1$. Thus, the minimal shift (\ref{shitf1Dfin}) determines the other Lamb shifts and increases with increasing of the parameter $m+f$.

\begin{figure}
\centering
\includegraphics[width=0.7\linewidth]{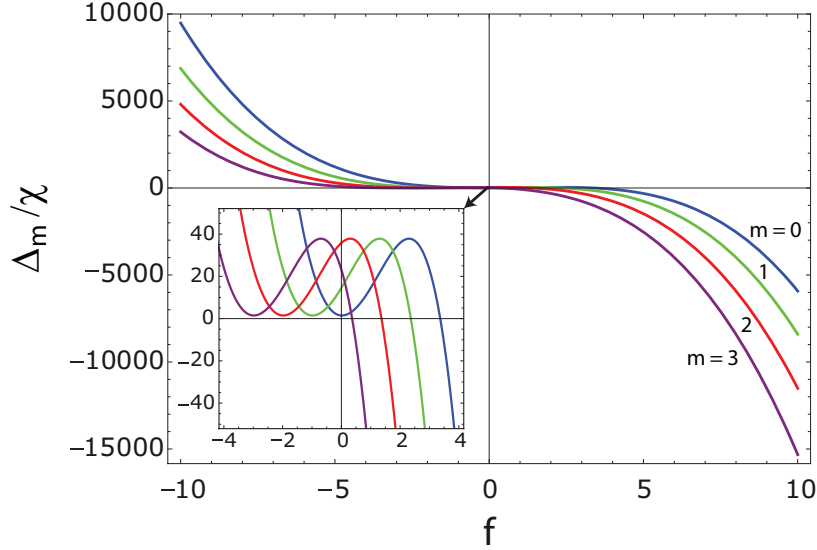}
\caption{Radiative shifts of energy levels with different angular quantum numbers $m=0,~1,~2,~3$, as a function of dimensionless magnetic flux $f=\Phi/\Phi_0$. The shifts are measured with respect to characteristic energy $\chi$.}
\label{Fig3}
\end{figure}

The behaviour of Aharonov-Bohm oscillations in the Lamb shift is demonstrated in Fig. \ref{Fig3}, where for convenience we introduced a characteristic energy $\chi = \alpha \varepsilon_{0}^2 /m_e$. We note that an exact value of a radiative shift strongly depends on the system parameters which define energy $\varepsilon_{0}$.
For $f=0$ the result (29) describes the radiative shifts of energetic levels $\Delta_{m}(0)$ of an electron in quantum ring without the magnetic field. In this case the minimal shift is realized for $m=0$ level.
We observe that the radiative shift dependence on the magnetic flux has maximum and minimum. In particular, the minima appear at the points $f=-m$. We find that Lamb shift is approximately equal to characteristic energy $\chi$ for small magnetic flux, but substantially grows for higher magnetic fields. Moreover, we find that the effect of vacuum corrections is larger for a quantum rings of smaller radius.

First, let us consider a semiconductor quantum ring of radius $R=20$ nm, with corresponding cut-off parameter being $k_{max} \approx 10$ meV. For the given parameters the characteristic energy is equal to $\chi = 0.13$ feV, and $\varepsilon_0 = 95~\mu$eV. Note, that for typical parameters of the semiconductor rings the values of the shifts are extremely small and lie in the range of nanoelectronvolts. 
However, one can expect that the shift will be dramatically increased in organic ring-shaped molecules, such as benzene, cycloalkanes and porphyrins, for which the value of the radius can become of the order of nanometers. Our estimation gives that for benzene ring of radius $R = 0.134$ nm the characteristic energy is $\chi = 0.065~\mu$eV. Thus, even for small values of piercing magnetic flux the Lamb shift riches microelectronvolt range. Moreover, considering large magnetic flux $f>5$ one gets sizeable shift of $10~\mu$eV order. 
For the porphyrin rings which have larger radius $R = 0.36$ nm \cite{OSullivan}, the characteristic energy can be estimated as $\chi = 1.3$ neV, leading to the Lamb shift which is order of magnitude smaller than in benzene case.

\subsection{Radiative shift for the  ring in a cavity}

For completeness, we now  shortly analyse the case of  quantum 1D ring pierced by the magnetic flux and  interacting with single mode in a cavity \cite{Kibis2011}. We calculate the shift   of $m$-energetic level due to emission and reabsorption of virtual photons of one-mode cavity by using the  $\vec{j}\vec{A}$ version of electron-radiation field interaction. 
We consider linearly polarized mode of cavity  in the Fock state with $N$ photons at the frequency $\omega$ with the  vector potential  given by
\begin{equation}
\vec{A}(t)=A_0 \hat{e}_x ( a e^{-i \omega t} + a^{\dagger} e^{i \omega t}).
\end{equation}

 By using the matrix element (\ref{matrix1}) the radiative  shift in the second order of the perturbation theory on interaction of the ring with cavity  photons is calculated as
\begin{equation}
\fl \Delta E_{m}= \frac{\alpha A_0^2 \varepsilon_{0}}{4 m_{e}} \Big\lbrace (N+1)\left( \frac{\Lambda_{+}^2}{\omega + \varepsilon_{0}\Lambda_{+}} + \frac{\Lambda_{-}^2}{\omega - \varepsilon_{0}\Lambda_{-}} \right) - N\left( \frac{\Lambda_{+}^2}{\omega - \varepsilon_{0}\Lambda_{+}} + \frac{\Lambda_{-}^2}{\omega + \varepsilon_{0}\Lambda_{-}} \right) \Big\rbrace,
\label{shift1Dcavity}
\end{equation}
where $\Lambda_{\pm}= 2(m+f)\pm 1$.

 Differently from the Lamb shift (\ref{Eshift}), Eq. (\ref{shift1Dcavity}) has not included both integration on virtual photon and renormalization procedure. Therefore the result  (\ref{Eshift}) does not contain small logarithmic factor  appearing for the Lamb shift.  Beside this the radiative shift for ring in the cavity can be controlled by the amplidude of field of cavity mode and increases with increasing of the amplitude$ A_0$. 

 The vacuum part for $N=0$ yields
\begin{equation}
\Delta E_{m}^{'}= \frac{\alpha A_0^2 \varepsilon_{0}}{4 m_{e}} \left( \frac{\Lambda_{+}^2}{\omega + \varepsilon_{0}\Lambda_{+}} + \frac{\Lambda_{-}^2}{\omega - \varepsilon_{0}\Lambda_{-}} \right),
\label{shift1Dcavity_0}
\end{equation}
with the minimal value 
\begin{equation}
\Delta E_{0}^{'}= \frac{\alpha A_0^2 \varepsilon_{0}}{2 m_{e}} \frac{1}{\omega + \varepsilon_{0}}
\label{shift1Dmin}
\end{equation}
that is realised for $m+f=0$.

\subsection{Radiative shift for 2D system}

In this section, we consider the Lamb shift of energy levels of electron confined in a 2D quantum ring. Using Eqs. (\ref{vtranz}), (\ref{Eshift}) one can obtain for the radiative shift of $E_{n,m}$ the following formula:
\begin{eqnarray}
\fl \Delta E_{n,m}=\frac{\alpha}{\pi}\sum_{n'}\Big[(E_{n',m+1}-E_{n,m})^{3}|R_{n',m+1;n,m}|^{2} \ln\frac{k_{max}}{|E_{n,m}-E_{n',m+1}|} \\ \nonumber +(E_{n',m-1}-E_{n,m})^{3} |R_{n',m-1;n,m}|^{2}\ln\frac{k_{max}}{|E_{n,m}-E_{n',m-1}|}\Big].
\label{Enm2D}
\end{eqnarray}

This expression contains the summation over all virtual transitions between electronic states due to emission and reabsorption of virtual photons. To calculate the value of shift, one needs to find the eigenfunctions and energy levels of the Hamiltonian $H_0$ (\ref{Ham2D}) for some specific confining potential $V(\rho)$. The eigenfunctions of $H_0$ are factorized: $\psi_{n,m}(\rho,\varphi)= R_{n,m}(\rho) \Phi_{m}(\varphi)$, and the radial wave function $R_{n,m}(\rho)$ is the solution of the following equation:
\begin{equation}
\label{RPHieq}
\fl -\frac{\hbar^2}{2m_e}\left( \frac{\partial^2}{\partial \rho^2} + \frac{1}{\rho} \frac{\partial}{\partial \rho} - \frac{m^2}{\rho^2} \right) R_{n,m}(\rho) + V(\rho)R_{n,m}(\rho) = \varepsilon(n,m) R_{n,m}(\rho).
\end{equation}
Here $n,~m$ are the principal and azimuthal quantum numbers, correspondingly, and the energy levels $E_{n,m}$ can be represented as a sum of azimuthal and radial parts, $E_{n,m}=\varepsilon_{m} + \varepsilon(n,m)$.

The Lamb shift physically appears due to the emission and reabsorption of the virtual photons. The corresponding correction to the energy in the case of 2D ring can be divided into two parts:
\begin{equation}
\Delta E_{n,m} = \Delta E_{n,m}^{(d)} + \Delta E_{n,m}^{(nd)}.
\end{equation}
The ``diagonal'' part $\Delta E_{n,m}^{(d)}$ contains the terms in which the absorption of the virtual photon does not change the principal number $n$, while azimuthal number changes by $\pm1$. Only these terms were accounted for the case of 1D ring. However, the rings of the final width include ``non-diagonal'' contribution to the Lamb shift given by $ \Delta E_{n,m}^{(nd)}$. In this term the absorption of virtual photon changes the value of $n$, and thus the term involves the summation over intermediate states $ \sum_{n'\neq n}(...)$.

Below we consider a narrow ring, without magnetic flux, in which width $d$ is much less than the radius of the ring, $d\ll R$. In this case the characteristic confinement energy of the radial motion $\varepsilon(n,m)$ is much greater than the spacing between azimuthal energy levels $\varepsilon_m$. Besides, in the case of a narrow ring one can assume that the radial wave functions $R_{n,m}$ depend only on the principal quantum number and energy spectrum can be represented as sum of azimuthal and radial contributions: $\varepsilon(n,m)=\varepsilon(n)+\varepsilon(m)$ . Indeed, the azimuthal motion enters into the equation for radial wavefunction in form of the additional centrifugal potential $\sim m^ 2/r^ 2$ which for the case of a narrow ring of the radius $R \gg d$ can be safely approximated by a constant value of $\sim m^2/R^2$. In this case the energy distances for diagonal transitions $E_{n,m}-E_{n,m\pm1}=\varepsilon(n,m)-\varepsilon(n,m\pm 1)$ are reduced to $\varepsilon_m - \varepsilon_{m\pm 1}=\varepsilon_0 \Lambda_{\pm}$,  and the diagonal matrix elements of transitions (\ref{Rnmnm}) can be rewritten as
\begin{equation}
R_{n,m';n,m} \simeq \int_{0}^{\infty} R_n(\rho)^2 \rho^2 d\rho=\langle R_n| \rho |R_n \rangle=\bar{R}_n.
\end{equation}
On the whole, we get the following expression for the ``diagonal'' part of the shift:
\begin{eqnarray}
\Delta E_{n,m}^{(d)}= \frac{\alpha}{\pi}\varepsilon_{0}^3 \Big[ \Lambda_{+}^3 \ln \left( \frac{k_{max}}{\varepsilon_0 |\Lambda_+|} \right) +  \Lambda_{-}^3 \ln\left( \frac{k_{max}}{\varepsilon_0 |\Lambda_-|} \right) \Big] \bar{R}_n.
\label{Enmd}
\end{eqnarray}

In the case of the narrow ring limit only the states with $n=0$ are occupied. To estimate the corresponding mean radius $\bar{R}_0$ we consider the model potential for a narrow ring in the form of displaced parabola,
\begin{equation}
V(\rho) = \frac{V_0}{2}(\rho - R)^2.
\label{parabolic}
\end{equation}
For $d \ll R$, we can use the harmonic approximation of  Eq. (\ref{RPHieq}) with potential (\ref{parabolic}) for the lowest eigenstate $R_0(\rho)$ and the energy $\varepsilon(0)$,
\begin{equation}
-\frac{\hbar^2}{2m_e} \frac{\partial^2 R_0(\rho)}{\partial \rho^2} + \frac{V_0}{2} (\rho - R)^2 R_0(\rho) = \varepsilon(0) R_0(\rho),
\label{ShrRad}
\end{equation}
for which
\begin{eqnarray}
\varepsilon_0=\frac{\hbar^2}{2 m_e d^2}=\frac{\hbar}{2}\sqrt{\frac{V_0}{m_e}},\\
R_0(\rho)=\left(\frac{2}{Rd\sqrt{\pi}} \right)^{1/2} e^{-(\rho - R)^2/2d^2}.
\label{R0rho}
\end{eqnarray}
This allows to get a simple result $\bar{R}_0 = R$. The value of the diagonal contribution to the Lamb shift thus coincides with earlier obtained result for purely 1D ring.

It should be mentioned that consideration of the simplest form of the potential (\ref{parabolic}) and the approximate solution (\ref{R0rho}) of the radial equation (\ref{ShrRad}) prove to be correct for estimation of the effective radius of ring. However, such approximation in principle is not fully correct. Note, also that in general 1D ring should be considered as the limit of a narrow 2D ring. However, this procedure can be only performed in details for each of the  given confined potential, but not in the general form. Some  examples have been described
 in \cite{Viefersa2004,Nowak2009}

 In the end of this section we provide the other approach that allows  to estimate the Lamb shift using Eq. (\ref{Enm2D}) in analogy to the standard  method of effective logarithm used in investigation of radiative shifts of atomic spectra \cite{Berestecki}.
 
Now let us consider the non-diagonal contribution to the Lamb shift [see Eqs. (35), (37) and also Eq. (44)] which involves the virtual radiative transitions between radial wave functions with different principal quantum numbers. We express the logarithms in Eq. (35)  as the logarithm depending from the characteristic mean energy $\bar{\varepsilon}$ of confined electron. In this approach the sum over non-diagonal transitions between radial wave function in Eq.(35) is transformed as
\begin{eqnarray}
\label{sum_m}
\fl \Delta E_{n,m}^{(nd)}=\frac{\alpha}{\pi}\sum_{n'\neq n}\bigg[(E_{n',m+1}-E_{n,m})^{3}|R_{n',m+1;n,m}|^{2}\ln\frac{k_{max}}{|E_{n,m}-E_{n',m+1}|} \\ \nonumber +(E_{n',m-1}-E_{n,m})^{3}|R_{n',m-1;n,m}|^{2}\ln\frac{k_{max}}{|E_{n,m}-E_{n',m-1}|}\bigg]\nonumber\\ \fl
=\frac{\alpha}{\pi}\sum_{n'\neq n}\Big[ (E_{n',m+1} - E_{n,m})^3 |R_{n',m+1;n,m}|^2 + (E_{n',m-1} - E_{n,m})^3 |R_{n',m-1;n,m}|^2 \Big] \ln(\frac{k_{max}}{\bar{\varepsilon}}).
\nonumber
\end{eqnarray}
Note, that in the narrow limit the matrix $ R_{n', m'; n, m}$ of nondiagonal  transitions with  $ n'\neq n$ are negligible and hence the nondiagonal part of the radiative shift is zero. Thus,  the radiative shift is only  expressed through the diagonal part in the limit of narrow 2D ring. 

We observe that the following standard formulas can be derived for the system described by the Hamiltonian (\ref{Ham2D})
\begin{eqnarray}
\fl S_{N}=\sum_{N'}(E_{N'}-E_{N})[\langle N|({\hat{e}^{*}}_{+}\vec{v})|N'\rangle\langle N'|(\hat{e}_{+}\vec{v})|N\rangle + \langle N|({\hat{e}^{*}}_{-}\vec{v})|N'\rangle\langle N'|(\hat{e}_{-}\vec{v})|N\rangle]\nonumber\\ =-\langle N|\left[(\hat{e}_{-}\vec{v}),\left[(\hat{e}_{+}\vec{v}),H_{0}\right]\right]|N\rangle,
\end{eqnarray}
where $|N\rangle = |n,m\rangle$, $|N'\rangle = |n',m'\rangle$. On the other hand
\begin{equation}
[(\hat{e}_{+}\vec{r}),H_0]=\frac{1}{m_e}(\hat{e}_{+}\vec{p})V(r),
\end{equation}
and then
\begin{equation}
[(\hat{e}_{-}\vec{r}),[(\hat{e}_{+}\vec{r}),H_0]]=\frac{1}{m_e^2}(\hat{e}_{-}\vec{p})(\hat{e}_{+}\vec{p})V(r).
\end{equation}
If the potential depends only on the radial coordinate, in the cylindric coordinates we obtain
\begin{eqnarray}
(\hat{e}_{-}\vec{v})(\hat{e}_{+}\vec{v})V=-\frac{\hbar^{2}}{{m_{e}}^{2}}
\left(\frac{\partial^{2}}{\partial\rho^{2}}+\frac{1}{\rho}\frac{\partial}{\partial\rho}\right)V(\rho) =-\frac{\hbar^2}{m_e^2} \nabla^2 V(\rho).
\end{eqnarray}
This expression is calculated by using Eq. (\ref{vtranz}) and involves the radial part of the Laplacian. The analogous formulas are well-known in atomic spectroscopy. Using Eqs. (35), (44)-(47) the radiative shifts of energy levels can be written approximately as
\begin{eqnarray}
\Delta E_{n,m}=\frac{\alpha}{\pi {m_{e}}^{2}}\ln\left(\frac{k_{max}}{\bar{\varepsilon}}\right)\int_{0}^{\infty}|R_{n,m}(\rho, \varphi)|^{2}\nabla^{2}V(\rho)
\rho d\rho d\varphi,
\end{eqnarray}
where $\bar{\varepsilon}$ is the characteristic energy.
This formula is expected to be useful for comparative analysis of vacuum radiative shifts of various ring energy levels. It can be also used for estimation of Lamb shifts for several model potentials. In particular, its application to the case of parabolic potential (40) immediately leads to the approximate value of the Lamb shift of lowest energy level
\begin{equation}
\Delta E_0 \simeq \frac{2 \alpha}{\pi m_e^2}V_0 \ln\left( \frac{k_{max}}{\bar{\varepsilon}} \right).
\end{equation}
We estimate this quantity for the system GaAs quantum ring embedded to AlGaAs substrate with parabolic potential and $V_0=1.68$ eV/nm$^2$. The estimation gives minimal value of $\Delta E_0 = 5.4$ neV, which largely overcomes the same quantity for 1D case. Next, assuming the porphyrin molecule placed on the metallic (e.g., aluminium) substrate \cite{Ishii,Wende} with parabolic confinement potential of $V_0=215$ eV/nm$^2$ the minimal radiative shift can be estimated as $\Delta E_0 \approx 0.5~\mu$eV.

\section{Conclusion}

We have studied radiative shifts of energy levels of an electron confined in a quantum ring. We have analyzed the non-relativistic part of the Lamb shift corresponding to the low-momentum spectral ranges of virtual photon for  both 1D and 2D models. It has been demonstrated that in the absence of the external magnetic field the minimal Lamb shift corresponds to the state of the minimal energy with $m=0$, which is qualitatively different from the case of the hydrogen atom where Lamb shifts are maximal for s-orbital states. Considering Aharonov-Bohm quantum ring pierced by a magnetic flux we demonstrate  flux-dependent oscillations in the vacuum self-energetic shift of the ground state. The low-frequency part of  the self-energy part calculated in this paper is the dominant radiative correction  to energetic levels of confined electron. Nevertheless, the  satisfactory consideration of the Lamb shift have to involve also the  high-frequency contribution, which will be the subject of future work.

The work was supported by FP7 IRSES projects POLAPHEN, POLATER, QOCaN, and ITN NOTEDEV. G. Yu. Kryuchkyan thanks the University of Iceland for hospitality and acknowledges support from the Armenian State
Committee of Science, the Project No.13-1C031.\\

\end{document}